\documentclass{elsart}
\usepackage{graphics,amssymb}
\journal{Physics Letters B}
\begin{document}

\begin{frontmatter}

\title{Generalizing the MOND description of rotation curves}
\author{Sandro S. e Costa},
\ead{sancosta@iagusp.usp.br} 
\author{R. Opher}
\ead{opher@orion.iagusp.usp.br}
\address{Instituto de Astronomia, Geof\'\i sica e Ci\^encias
Atmosf\'ericas -- USP\\ Av. Miguel St\'efano, 4200 - CEP 04301-904 - S\~ao Paulo - SP - Brazil}

\begin{abstract}
We present new mathematical alternatives for explaining rotation
curves of spiral galaxies in the MOND context. For given total masses, it is
shown that various mathematical alternatives to MOND, while predicting
flat rotation curves for large radii ($r/r_d\gg 4$, where $r_d$ is the characteristic radius of the galactic disc), predict curves with different peculiar features for smaller radii ($0.1<r/r_d\lesssim 4$). They are thus testable against observational data.
\end{abstract}


\end{frontmatter}

\section{Introduction}

The first mathematical descriptions of the effects of gravity, made by
Galileo in his study of the free fall of bodies and by Kepler in his study
of planetary motions, were purely empirical. Though Newton offered a coherent
explanation of what was behind the laws governing gravitational effects, it was only with Einstein's
General Relativity that we had an apparently complete theory of gravity.

However, at the end of the 20$^{th}$ century, a new enigma concerning the
motion of `celestial bodies' emerged, in particular, in studying rotation
curves of spiral galaxies. While Newton's law of gravity predicts that the
velocity of rotation in the interior of a galaxy should fall with increasing
distance from the galactic center if the observed light traces mass,
what is observed is the maintenance of a constant velocity with increasing
radius, generating flat rotation curves \cite{Peebles}.

Two simple ways of dealing with this problem have been suggested:
\begin{enumerate}
\item assuming that there is more mass ({\it i.e.}, dark matter) in galaxies than is observed;
\item modifying the law of gravity.
\end{enumerate}

While much work has been done in the search for possible particle candidates for
dark matter \cite{DM1}, very little has been done to explore the possibilities of
modified gravity laws. Until now, the most popular suggestion for a modified 
gravitational law has been Modified Newtonian Dynamics, or, MOND 
\cite{Milgrom1,Milgrom2,Milgrom3}. In MOND the acceleration $a$ of
a body in an external gravitational field is not exactly equal to the
acceleration $g_N$ obtained from the Newtonian gravitational force.
Mathematically, one can write $a\mu =g_N$, where $\mu \left( x\right) $ is a
dimensionless function of the ratio $x\equiv a/a_0$ of the acceleration $a$
to an empirically determined constant $a_0$. Only in the limit $a\gg a_0$ is
Newtonian gravity restored. The strongest objection to MOND is that it
does not have a relativistic theory supporting it.
For recent articles criticizing MOND, see Scott {\it et al.} (2001) \cite
{Fin} and Aguirre {\it et al.} (2001) \cite{LyAlpha}. For a recent positive
review of MOND, see Sanders (2001) \cite{Sanders}.

The objective of this letter is to expand the original MOND proposal
by presenting mathematical alternatives for
the modified gravitational law. Specifically, we present several alternative
mathematical alternative formulations for the dimensionless function $\mu $, thus following closer
the structure of the pioneering work of MOND by Milgrom \cite
{Milgrom1,Milgrom2,Milgrom3}. In the next section we
present the basics of MOND. Simulated rotation curves for several possible
MONDian-like functions are given in Section \ref{sec:formulas}. The
final section presents some brief conclusions and perspectives for future
work.

\section{MOND}

As discussed in the introduction, the original MOND proposal uses the
relation
\begin{equation}
\label{MOND}a\mu =g_N\;\;, 
\end{equation}
where $g_N$ is the usual Newtonian acceleration and $\mu\left( x\right)$ is
a function which obeys
\begin{equation}
\mu\left( x\right)=\left\{ 
\begin{array}{lll}
1 & & \left( x\gg 1\right) \\ 
x & & \left( x\ll 1\right) 
\end{array}
\right.\;\;. 
\end{equation}
Therefore, in the limit of large accelerations, $x\gg 1$, the
usual Newtonian gravity law is obtained. In
the other extreme, $x\ll 1$, however, we have 
\begin{equation}
a\mu \simeq a^2/a_0=g_N=GMr^{-2}\;\;. 
\end{equation}
Thus, using $a=V^2r^{-1}$, where $V$ is the rotation velocity of the galaxy, 
\begin{equation}
V\simeq \left( GMa_0\right) ^{1/4}\;\;, 
\end{equation}
which is a constant, as is observed for large galactic radii.

It is common in the literature ({\it e.g.} \cite{Fin}, \cite{LyAlpha}) to
use the expression 
\begin{equation}
\label{mu}\mu\left( x\right)=x\left( 1+x^2\right) ^{-1/2} \;\;. 
\end{equation}
This formula, proposed by Milgrom \cite{Milgrom1,Milgrom2,Milgrom3}, has
the advantage of being invertible. With it one can solve eq. (\ref{MOND}) analytically
for the acceleration $a$ and, consequently, for the
rotation velocity $V$ as a function of the radius $r$. However, other functions are also
possible, and are discussed in the next
section.

\section{\label{sec:formulas}Alternative mathematical formulations of MOND}

In his work on the implications of MOND for galaxies \cite{Milgrom2}%
, Milgrom used as a model for a spiral galaxy of total mass $M$, a disc of mass $M_d$
and a central spheroidal bulge of mass $M_s$. The fractional masses for the disc and the spherical bulge are $%
\alpha _d\equiv M_d/M$ and $\alpha _s\equiv M_s/M=1-\alpha _d$, respectively, so that the
total fractional mass $\gamma \equiv M\left( r\right) /M$ inside a radius $%
r\equiv sr_d$ is 
\begin{equation}
\label{gamma}\gamma \left( s\right) =\alpha _d\gamma _d\left( s\right)
+\left( 1-\alpha _d\right) \gamma _s\left( s\right) \;\;, 
\end{equation}
where \cite{Milgrom2}
\begin{equation}
\label{gammad}\gamma _d\left( s\right) =\left( \frac{s^3}2\right) \left[
I_0\left( \frac s2\right) K_0\left( \frac s2\right) -I_1\left( \frac
s2\right) K_1\left( \frac s2\right) \right] \;, 
\end{equation}
\begin{equation}
\label{gamas}\gamma _s\left( s\right) =\frac{\sqrt{8\pi ^3}b^{-9}}{m_t}%
\gamma \left( 8.5,bs^{1/4}u^{-1/4}\right) \;\;, 
\end{equation}
and $\gamma \left( a,z\right) =\int_0^ze^{-t}t^{a-1}dt$ is the
incomplete gamma function. $b=7.66924944$ and $m_t=2.4082\times 10^{-3}$
are numerical constants. The dimensionless variable $s=r/r_d=ur/r_s$ is
the ratio of the radius $r$ to the characteristic length $r_d$. 
The ratio of $r_s$ to $r_d$, $u=r_s/r_d$, is less than unity. 
The radii $r_d$ and $r_s$ are obtained, in practice, by
adjusting the luminosity profiles of the spheroidal and disc components, 
using the empirical law of de Vaucoulers for the spherical bulge and an 
exponential function for the disc.

\begin{figure}[p]
\centerline{\scalebox{.6}{\includegraphics{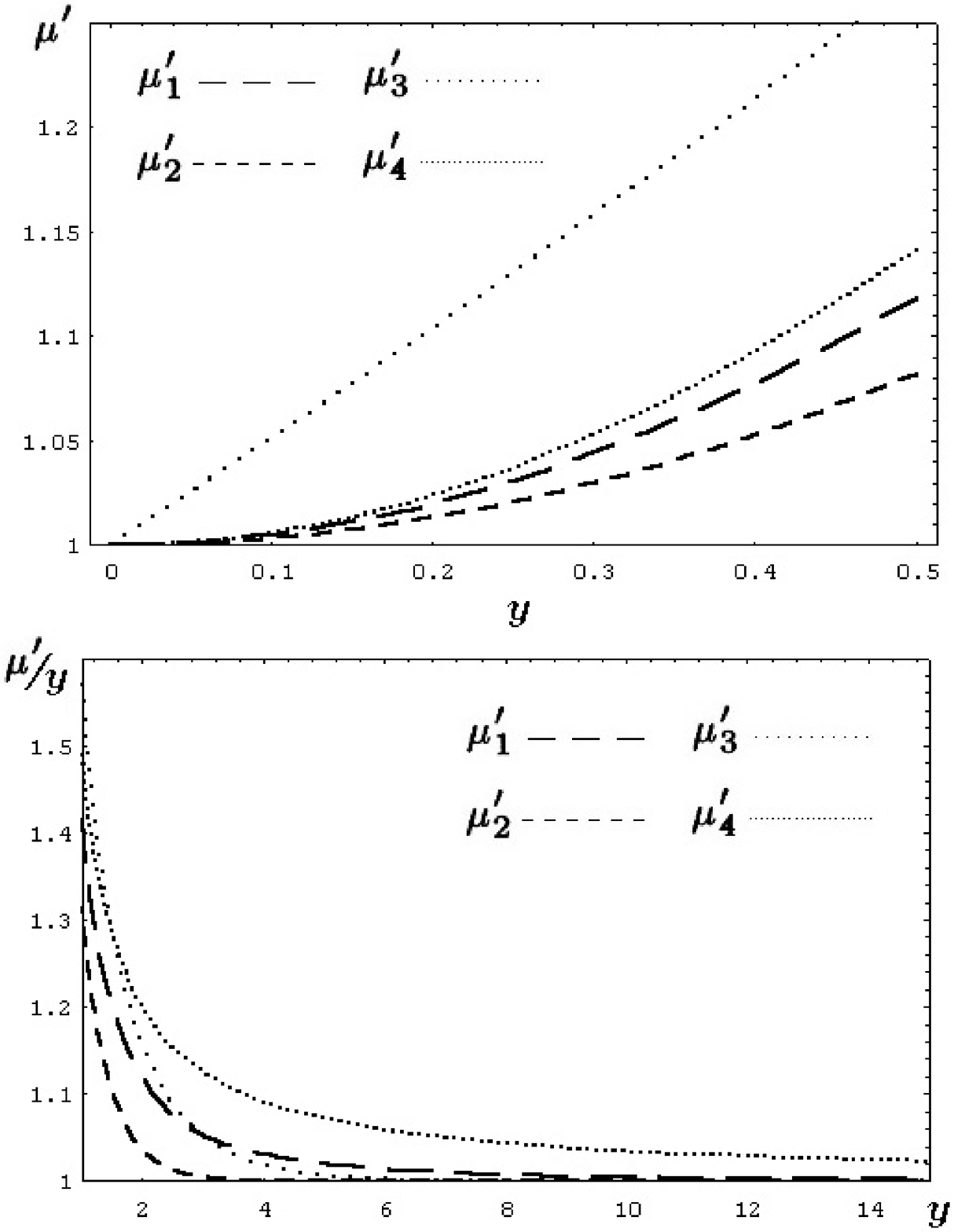}}}
\caption{\label{galmond4}
Top: Curves showing the behaviour of the 
$\mu ^{\prime }$ functions as a 
function of $y\equiv r\sqrt{a_0G^{-1}M^{-1}\left( r\right)}$
for $y<1$.
Bottom: Curves showing the behaviour of $\mu ^{\prime }/y$ as a
function of $y$ 
for $y>1$ 
(the functions $\mu _1^{\prime }$, $\mu _2^{\prime }$, 
$\mu _3^{\prime }$ and $\mu _4^{\prime }$ versus $y$ are 
defined in eq. (\protect{\ref{functions}})).} 
\end{figure}

Following the MOND proposal, we define 
\begin{equation}
\label{MOND2}a=\mu ^{\prime }\left( y\right) g_N\;, 
\end{equation}
where $\mu ^{\prime }$ is a dimensionless 
function with a
dimensionless argument $y\equiv r \sqrt{a_0G^{-1}M^{-1}(r)}$, similar to the $\mu $ of Milgrom 
\cite{Milgrom1,Milgrom2,Milgrom3} in eq. (\ref{mu}). This new
function $\mu ^{\prime }\left( y\right) $ is such that 
\begin{equation}
\label{constraints}\mu ^\prime\left( y\right)=\left\{ 
\begin{array}{lll}
y & & \left( y\gg 1\right) \\ 
1 & & \left( y\ll 1\right) 
\end{array}
\right.\;\;. 
\end{equation}
We investigate the following functions $\mu ^{\prime }\left( y\right)$ which obey 
the constraints of eq. (\ref{constraints}): 
\begin{equation}
\label{functions}\left\{ 
\begin{array}{l}
\mu ^\prime _1=
\sqrt{1+y^2} \\ \mu ^\prime _2=y\coth y \\ 
\mu ^\prime _3=y\left( 1-e^{-y}\right) ^{-1} \\ 
\mu ^\prime _4=y\left( \coth 3y-\frac 1{3y}\right) ^{-1} 
\end{array}
\right. \;\;. 
\end{equation}
The behaviour of each of these functions as a function of $y$ 
can be seen in the expansions \cite{GR}%
\begin{equation}
\left\{ 
\begin{array}{l}
\begin{array}{ll}
\mu _1^{\prime } & =\sum_{k=0}^\infty \Gamma \left( n-1/2\right) \left( 2
\sqrt{\pi }n!\right) ^{-1}(-1)^{n-1}y^{2n} \\  & 
=1+y^2/2-y^4/8+y^4/16-5y^6/128+...
\end{array}
\\ 
\begin{array}{llr}
\mu _2^{\prime } & =1+2y^2\sum_{k=1}^\infty \left( y^2+k^2\pi ^2\right)
^{-1} &  \\  
& \simeq 1+y^2/3-y^4/45+2y^6/245+... & \left( y<\pi\right) 
\end{array}
\\ 
\begin{array}{llr}
\mu _3^{\prime } & =y\sum_{k=0}^\infty e^{-ky} &  \\  
& \simeq 1+y/2+y^2/12-y^4/720+y^6/30240+... & \left( y<2\pi \right)
\end{array}
\\ 
\begin{array}{llr}
\mu _4^{\prime }& =\frac 16\left[ \sum_{k=1}^\infty \left( 9y^2+k^2\pi
^2\right) ^{-1}\right] ^{-1} & \\
& \simeq 1+27y^2/45-\left( 27y/45\right) ^4+\left( 27y/45\right) ^6+... & \left( y\ll 1\right)
\end{array}
\end{array}
\right.\;\;.  
\end{equation}
The functions are plotted in Figure \ref{galmond4}. 

\begin{figure}[p]
\centerline{\scalebox{.6}{\includegraphics{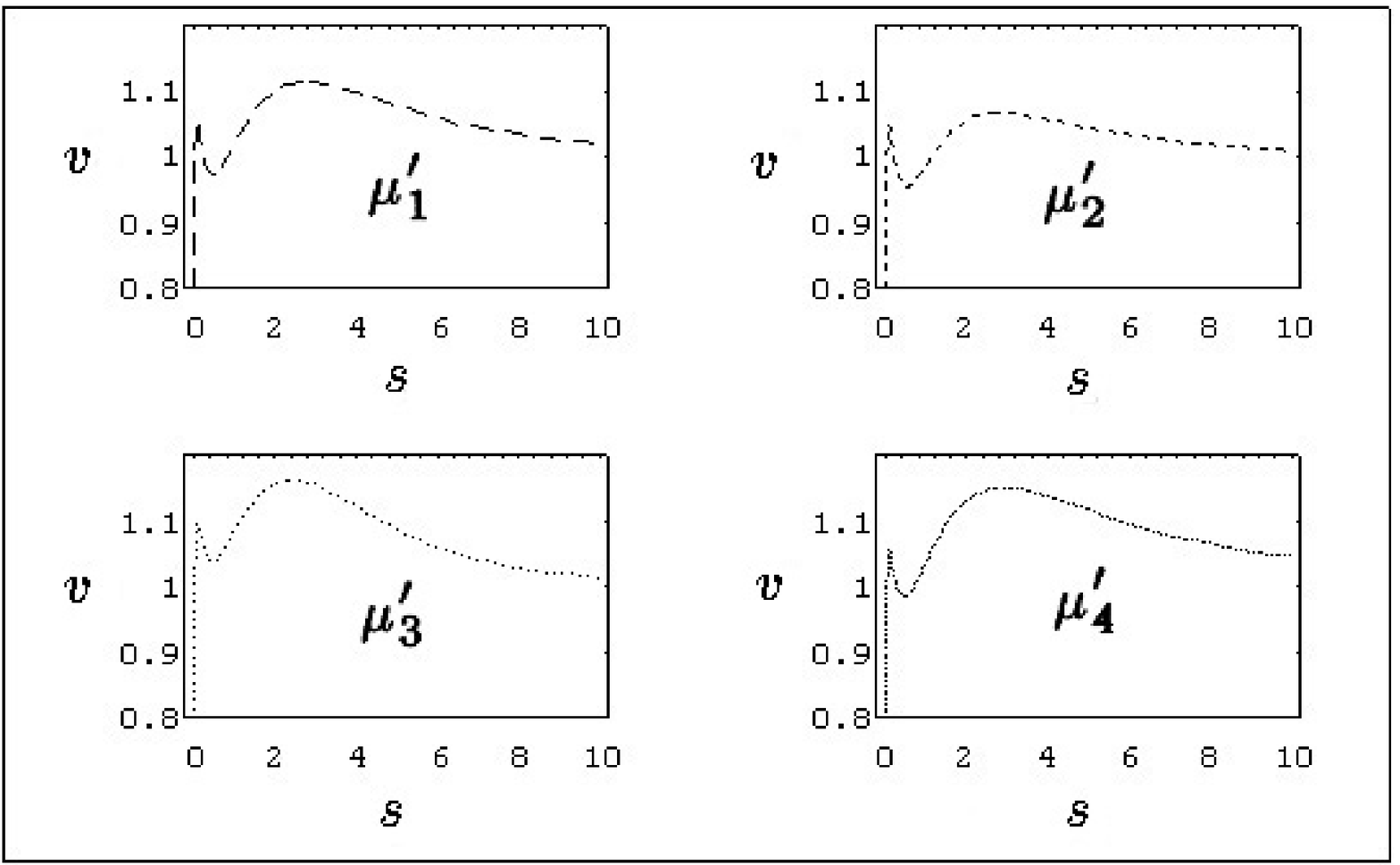}}}
\caption{\label{galmond6}
Curves of $v\equiv V\left( r\right) /\left( GMa_0\right) ^{1/4}$ 
as a function of $s\left( =r/r_d\right)$ and $\mu ^{\prime }$. 
The functions $\mu ^{\prime }$ are defined in eq. (\protect{\ref{functions}}). 
In all graphs, $\alpha _d=3/4$, $u=1/4$, and $M=M_0$, where $M_0$ is an arbitrary mass. 
The functions $\alpha _d$ and $u$ are defined as $\alpha _d\equiv M_d/M$ 
and $u\equiv r_s/r_d$, where $M_d$ is the mass of 
the disc and $r_s$ ($r_d$) is the effective radius of the spherical bulge (disc).}
\end{figure}

\begin{figure}[p]
\centerline{\scalebox{.6}{\includegraphics{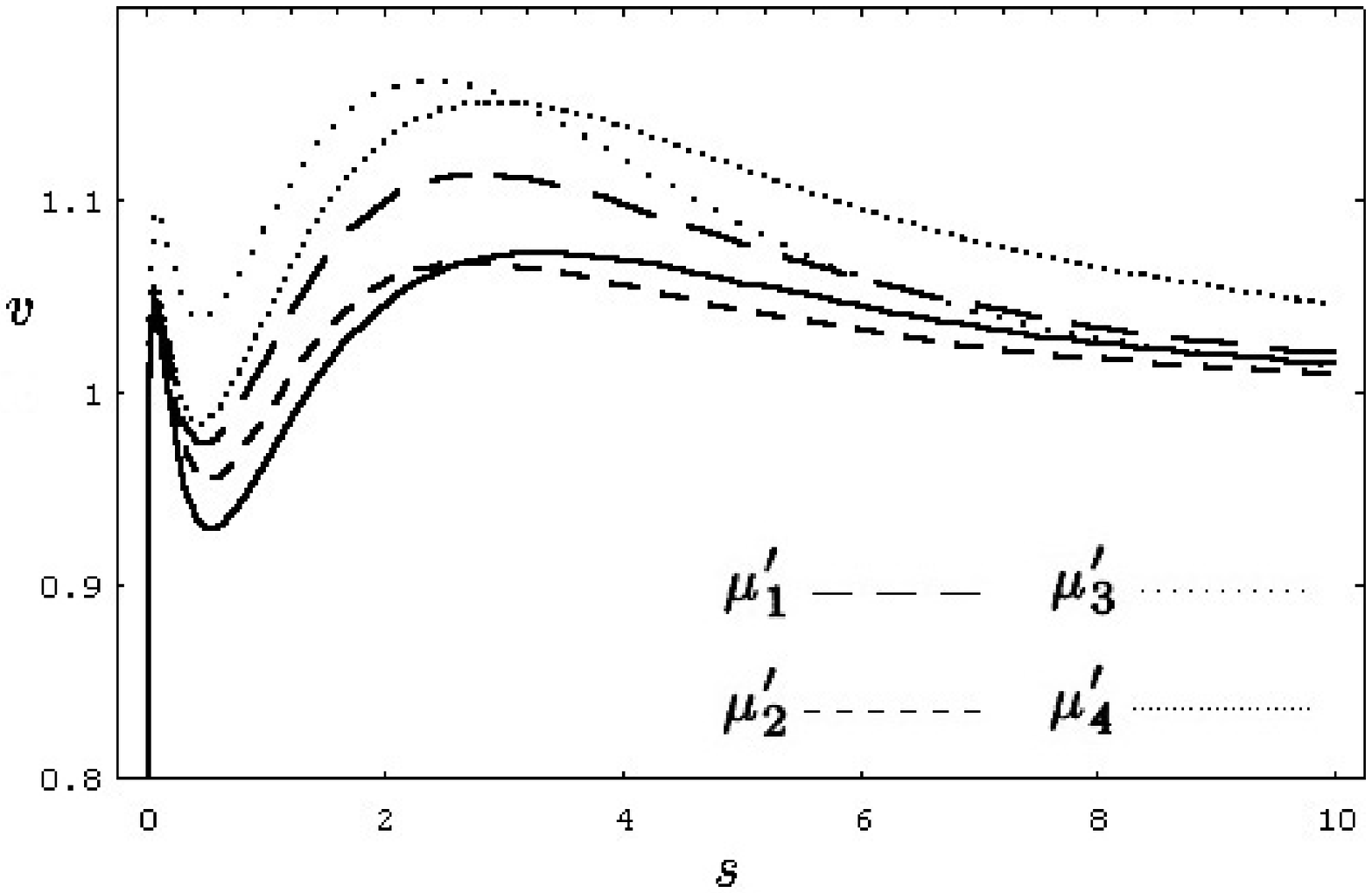}}}
\caption{\label{galmond7} 
Curves of $v\equiv V\left( r\right) /\left( GMa_0\right) ^{1/4}$ 
as a function of $s\left( =r/r_d\right)$ and $\mu ^{\prime }$. 
The functions $\mu ^{\prime }$ are defined in eq. (\protect{\ref{functions}}). 
The solid line is the curve obtained using Milgrom's original proposal, 
eq. (\protect{\ref{mu}}). For all curves $M=M_0$, where $M_0$ is an arbitrary mass, $\alpha _d=3/4$,
$u=1/4$. The functions $\alpha _d$ and $u$ are defined as
$\alpha _d\equiv M_d/M$
and $u\equiv r_s/r_d$, where $M_d$ is the mass of the disc and $r_s$ ($r_d$) 
is the effective radius of the spherical bulge (disc).}
\end{figure}

Using these functions, together with equations (\ref{gamma}), (\ref{gammad}) 
and (\ref{gamas}), we obtain curves for the
dimensionless rotation velocity $v\equiv V\left( r\right) /\left( GMa_0\right) ^{1/4}$ 
as a function of $s=r/r_d$ for different values of $M$, $\alpha _d=M_d/M$, 
and $u=r_s/r_d$. The
curves are shown in Figures \ref{galmond6} and \ref{galmond7}.

\section{Conclusion}

Inspection of Figures \ref{galmond6} and \ref{galmond7}
shows clearly that all the functions $\mu ^{\prime }\left( y\right) $ produce flat
rotation curves. This is true not only for the particular values of $M$, $%
\alpha _d$, and $u$ of the figures, but for the entire range of physically
reasonable values for these parameters. Figure \ref{galmond7}
shows that a comparison between the curves obtained, using the different $\mu
^{\prime }$ functions presented, together with the original Milgrom proposal
(eq. (\ref{mu})), may be useful to distinguish between them, since each curve has a peculiar feature in the region $0.1<r/r_d\lesssim 4$.

It would be interesting to test the formulas presented here against
observational data, noting that $\alpha _d$ and $u$ are
not free parameters, but are given by the luminosity profiles of the galaxies.
The mass $M$ and the constant $a_0$ are the only free parameters to be
adjusted. The study of different galaxies gives a single value for $a_0$, the 
mass $M$ and the mass-luminosity ratio, $M/L$, of each galaxy.
$\mu $ and $\mu ^\prime $ can lead to
different relativistic extensions of MOND, important
for future studies. For instance, using the expression for the gravitational potential $\varphi \left( r\right) =-\int a\left( r\right) dr$, valid for purely radial forces, one can naively ascribe a $\varphi\left( r\right)$ to the modified gravitational laws obtained with $\mu ^\prime_1$, $\mu ^\prime_2$ and $\mu ^\prime_3$, for example, 
\begin{equation}
\varphi _{\mu ^{\prime }_1}\left( r\right) =-\frac{GM}{r}\sqrt{1+\frac{r^2a_0}{GM}}+ \sqrt{GMa_0}\mathrm{arcsinh}\left( r\sqrt{\frac{a_0}{GM}}\right) +\varphi_0 \;\;, 
\end{equation}
\begin{equation}
\varphi _{\mu ^{\prime }_2}\left( r\right) =  -\frac{GM}{r} +2\sqrt{GMa_0}\sum_{n=1}^{\infty }\frac{1}{n\pi }\arctan\left(\frac
{r}{n\pi }\sqrt{\frac{a_0}{GM}}\right) +\varphi_0  \;\;, 
\end{equation}
and 
\begin{equation}
\varphi _{\mu ^{\prime }_3}\left( r\right) =\sqrt{GMa_0}\left[\ln \left( r\sqrt{\frac{a_0}{GM}}\right) +\sum_{n=1}^{\infty }{\mathrm Ei}\left( -nr\sqrt{\frac{a_0}{GM}}\right)\right] +\varphi_0
\;\;,
\end{equation}
where $\varphi_0$ is a constant of integration.

Therefore, in the search for a complete theory for MOND, it is
important to study alternative MONDian functions. The
MONDian functions given in this letter can be seen as a step in
this direction.

\section*{Acknowledgements}

SSC thanks the Brazilian agency FAPESP for financial support under grant
00/13762-6. Both SSC and RO thank FAPESP for partial support under grant
00/06770-2 and the Brazilian project PRONEX/FINEP (41.96.0908.00) for
partial support.

\end{document}